# Nodal superconductivity and superconducting domes in the topological Kagome metal CsV$_3$Sb$_5$

C. C. Zhao[1,6], L. S. Wang[1,6], W. Xia[2,3,6], Q. W. Yin[4,6], J. M. Ni[1], Y. Y. Huang[1], C. P. Tu[1], Z. C. Tao[2], Z. J. Tu[4], C. S. Gong[4], H. C. Lei[4,+], Y. F. Guo[2,‡], X. F. Yang[1,#], and S. Y. Li[1,5*]

[1] State Key Laboratory of Surface Physics, Department of Physics, Fudan University, Shanghai 200438, China.

[2] School of Physical Science and Technology, ShanghaiTech University, Shanghai 201210, China.

[3] ShanghaiTech Laboratory for Topological Physics, Shanghai 201210, China.

[4] Department of Physics and Beijing Key Laboratory of Opto-electronic Functional Materials and Micro-nano Devices, Renmin University of China, Beijing 100872, China.

[5] Shanghai Research Center for Quantum Sciences, Shanghai 201315, China.

[6] These authors contributed equally.

Corresponding author. Email: hlei@ruc.edu.cn (H.C.L); guoyf@shanghaitech.edu.cn(Y.F.G); yangxiaofan@fudan.edu.cn(X.F.Y.); shiyan_li@fudan.edu.cn (S.Y.L.)

## Abstract

Recently superconductivity was discovered in the Kagome metal $A$V$_3$Sb$_5$ ($A$ = K, Rb, and Cs), which has an ideal Kagome lattice of vanadium. These V-based superconductors also host charge density wave (CDW) and topological nontrivial band structure. Here we report the ultralow-temperature thermal conductivity and high pressure resistance measurements on CsV$_3$Sb$_5$ with $T_c \approx 2.5$ K, the highest among $A$V$_3$Sb$_5$. A finite residual linear term of thermal conductivity at zero magnetic field and its rapid increase in fields suggest nodal superconductivity. By applying pressure, the $T_c$ of CsV$_3$Sb$_5$ increases first, then decreases to lower than 0.3 K at 11.4 GPa, showing a clear first superconducting dome peaked around 0.8 GPa. Above 11.4 GPa, superconductivity re-emerges, suggesting a second superconducting dome. Both nodal superconductivity and superconducting domes point to unconventional superconductivity in this

**V-based superconductor. While our finding of nodal superconductivity puts a strong constrain on the pairing state of the first dome, which should be related to the CDW instability, the superconductivity of the second dome may present another exotic pairing state in this ideal Kagome lattice of vanadium.**

## Introduction

Finding unconventional superconductors and understanding their superconducting mechanism is the frontier of condensed matter physics, e.g., the heavy-fermion superconductors, Cu-based and Fe-based high-temperature superconductors[1], and recent Ni-based superconductors[2]. Unlike the conventional *s*-wave superconductors, the wave function of Cooper pairs for unconventional superconductors is usually not *s*-wave. Symmetry imposed nodes (gap zeros) are often observed, such as in *d*-wave cuprate superconductors and heavy-fermion superconductor $CeCoIn_5$ (ref. [3,4]). Note that the Fe-based superconductors are exceptions, in which both multiple *s*-wave gaps and nodal gaps are found[5]. Furthermore, the superconducting pairing mechanism of unconventional superconductors is not phonon-mediated. This usually manifests as a superconducting dome neighbouring a magnetic order or density-wave order in the phase diagram, and spin or density-wave fluctuations are considered as the major pairing glue[1]. The superconducting gap structure and superconducting dome nearby an ordered state provide important clues to the underlying pairing mechanism.

Recently, superconductivity was discovered in a new family of V-based compounds $AV_3Sb_5$ ($A$ = K, Rb, and Cs)[6-9]. These compounds have the ideal Kagome lattice of vanadium coordinated by antimony, with the $A$ atoms intercalated between the layers, as seen in Fig. 1a and 1b. The superconducting transition temperature $T_c$ is 0.93, 0.92, and 2.5 K for $A$ = K, Rb, and Cs, respectively[7-9]. While there are no local magnetic moments[10], these V-based superconductors manifest a charge density wave (CDW) order at 78, 103, and 94 K, respectively[6-9]. Interestingly, for $KV_3Sb_5$, high-resolution scanning tunneling microscopy (STM) study demonstrated that such charge order in the frustrated Kagome lattice is topological[11], which leads to a giant anomalous Hall effect[12], and can also be a strong precursor of unconventional superconductivity[11]. Moreover, topological nontrivial band structures, including multiple Dirac points and possible surface state, were revealed by angle-resolved photoemission spectroscopy (ARPES) measurements combined with density-functional theory (DFT) calculations[7,12]. In this context, the Kagome metal $A$V$_3$Sb$_5$ provides a great platform to study the interplay of superconductivity, CDW, frustration, and topology. It will be very important to understand the superconducting state first.

In this paper, we present ultralow-temperature thermal conductivity measurements of CsV$_3$Sb$_5$ single crystal to investigate its superconducting gap structure. The data in zero and magnetic fields clearly demonstrate that there are nodes in the superconducting gap. Furthermore, two superconducting domes in the temperature-pressure phase diagram are revealed by resistance measurement under pressures up to 47.0 GPa. These results suggest unconventional superconductivity in CsV$_3$Sb$_5$. We discuss the possible novel superconducting states in these V-based superconductors.

## Results and discussion

Figure 1c plots the XRD pattern of CsV$_3$Sb$_5$ single crystal, showing that the largest natural face is (00$l$) plane. In Fig. 1d, the magnetization measured in 1 T with zero-field and field cooling modes displays a sharp drop at 94 K, which is the CDW transition reported previously[7]. The low-temperature magnetic susceptibility measured at 10 Oe with zero-field and field cooling modes is plotted in Fig. 2a. The onset of the superconductivity is at 2.5 K, which is also consistent with previous report[7].

In Fig. 2b, we present the temperature dependence of in-plane resistivity for CsV$_3$Sb$_5$ single crystal in magnetic fields up to 2 T. In zero field, the $T_c$ defined at 10% drop of normal-state resistivity ($T_c^{10\%}$) and zero resistivity $T_c^{zero}$) are 3.6 K and 2.7 K, respectively. The $T_c^{zero}$ is slightly higher than the onset $T_c$ from magnetic susceptibility measurement. The normal-state resistivity shows a very weak temperature dependence below 6 K. A simple extrapolation gives residual resistivity $\rho_0$(0T) = 4.75 $\mu\Omega$ cm and $\rho_0$(0.2T) = 4.82 $\mu\Omega$ cm, respectively. The temperature dependence of $H_{c2}$, determined by the $T_c^{zero}$ values in Fig. 2b, is plotted in Fig. 2c. The red line is a linear fit to $\mu_0H_{c2}(T)$, and $\mu_0H_{c2}(0) \approx 0.47$ T is roughly estimated.

The ultralow-temperature heat transport measurement is an established bulk technique to probe the superconducting gap structure[13]. Figure 3 presents the in-plane thermal conductivity results of CsV$_3$Sb$_5$ single crystal. At very low temperatures, the thermal conductivity can usually be fitted to $\kappa/T = a + bT^{\alpha-1}$ (ref. [14,15]), in which the two terms $aT$ and $bT^\alpha$ represent contributions from electrons and phonons, respectively. The power $\alpha$ is typically between 2 and 3, due to specular reflections of phonons at the boundary[14,15]. In zero field, the fitting of the data below 0.5 K gives a finite residual linear term $\kappa_0/T \equiv a = 0.22 \pm 0.04$ mW K$^{-2}$ cm$^{-1}$ and $\alpha = 2.70 \pm 0.12$. Figure 3a is plotted as $\kappa/T$ vs $T^{1.70}$ to show the data more clearly.

For *s*-wave nodeless superconductors, there are no fermionic quasiparticles to conduct heat as $T \rightarrow 0$, since the Fermi surface is entirely gapped. Therefore, there is no residual linear term $\kappa_0/T$, as seen in InBi and NbSe$_2$ (ref. [16,17]). However, for nodal superconductors, a substantial $\kappa_0/T$ in zero field contributed by the nodal quasiparticles can be found[13]. For example, $\kappa_0/T$ of the overdoped ($T_c = 15$ K) *d*-wave cuprate superconductor Tl$_2$Ba$_2$CuO$_{6+\delta}$ (Tl-2201) is 1.41 mW K$^{-2}$ cm$^{-1}$, which is about 36% of the normal-state value $\kappa_{N0}/T$ (ref. [18]), and $\kappa_0/T = 17$ mW K$^{-2}$ cm$^{-1}$ for Sr$_2$RuO$_4$, which is about 9% of its $\kappa_{N0}/T$ (ref. [19]). For clean Fe-based superconductor KFe$_2$As$_2$, $\kappa_0/T = 3.8$ mW K$^{-2}$ cm$^{-1}$ is about 3.2% of the normal-state value $\kappa_{N0}/T$ (ref. [20]). For ultraclean YBa$_2$Cu$_3$O$_7$ (i.e., very low $\rho_0$ and very high $\kappa_{N0}/T$ in the normal state), $\kappa_0/T = 0.16$ mW K$^{-2}$ cm$^{-1}$ is a universal value for *d*-wave superconducting gap[21]. Therefore, the finite $\kappa_0/T = 0.22$ mW K$^{-2}$ cm$^{-1}$ of CsV$_3$Sb$_5$ in zero field is a strong evidence for the presence of nodes in the superconducting gap[13].

Further information on the superconducting gap structure can be obtained by examining the behavior of field-dependent $\kappa_0/T$ (ref. [13]). In Fig. 3b, the thermal conductivity of CsV$_3$Sb$_5$ in magnetic fields up to 0.25 T are plotted. By applying a very small field 0.01 T, one can see a large enhancement of thermal conductivity. Since all the curves in magnetic fields are roughly linear, we fix $\alpha$ to 2, thus fit the data to $\kappa/T = a + bT$ and obtain the $\kappa_0/T$ for each magnetic field. For $\mu_0H = 0.16$ T, the fitting gives $\kappa_0/T = 4.15 \pm 0.13$ mW K$^{-2}$ cm$^{-1}$. Further increasing field to 0.25 T does not increase the thermal conductivity, therefore we take 0.16 T as the bulk $H_{c2}(0)$. Note that this value is lower than that obtained from resistivity measurements. Interestingly, there is about 20% violation of Wiedemann-Franz law in the normal state, when we compare the $\kappa_0/T$ at 0.16 T to its Wiedemann-Franz law expectation $L_0/\rho_0$ (0.16T) $\approx 5.08$ mW K$^{-2}$ cm$^{-1}$, with the Lorenz number $L_0 = 2.45 \times 10^{-8}$ W $\Omega$ K$^{-2}$ and $\rho_0$(0.16T) $\approx \rho_0$(0.20T) = 4.82 $\mu\Omega$ cm. The origin of this violation is not clear to us at this stage. It may come from the topological nontrivial band structure, i.e., either the Dirac quasiparticles violate the Wiedemann-Franz law, or the surface state conducts electrical current much better than the heat current. We leave this issue for future investigation.

The normalized values of $[\kappa_0/T]/[\kappa_{N0}/T]$ as a function of $H/H_{c2}$ for CsV$_3$Sb$_5$ is plotted in Fig. 3c, with $\kappa_{N0}/T = 4.15$ mW K$^{-2}$ cm$^{-1}$ and $H_{c2} = 0.16$ T. For comparison, similar data of the clean *s*-wave superconductor Nb (ref. [22]), the dirty *s*-wave superconducting alloy InBi (ref. [16]), the multiband *s*-wave superconductor NbSe$_2$ (ref. [17]), and an overdoped *d*-wave cuprate superconductor Tl-2201 (ref. [18]), are also plotted. For CsV$_3$Sb$_5$, the field dependence of $\kappa_0/T$ clearly mimics the behaviour of Tl-2201. The rapid increase of $\kappa_0/T$ in magnetic field should come from the Volovik effect of nodal quasiparticles, thus provides further evidence for nodes in the superconducting gap[13]. To our knowledge, so far all nodal superconductors have unconventional pairing mechanism[1]. In this regard, the nodal gap we demonstrate from thermal conductivity results suggests unconventional superconductivity in CsV$_3$Sb$_5$.

To get further clue to the pairing mechanism in CsV$_3$Sb$_5$, we map out its temperature-pressure phase diagram by resistance measurement under pressures. Figure 4a presents the low-temperature resistance of CsV$_3$Sb$_5$ single crystal under various pressures up to 11.4 GPa. At ambient pressure, the $T_c^{10\%}$ is 3.6 K. With increasing pressure, the $T_c^{10\%}$ first increases sharply to 5.6 K at 0.8 GPa, enhanced by 56%. Under this pressure, applying magnetic field gradually suppresses the superconducting transition, as shown in Fig. 4b. With further increasing pressure, $T_c^{10\%}$ decreases slowly to below 0.3 K at 11.4 GPa. The non-monotonic pressure dependence of $T_c^{10\%}$ is plotted in Fig. 4c, which shows a clear superconducting dome. Since the CDW usually competes with superconductivity, it is expected that the CDW order may be suppressed near the optimal pressure $p_c$ ~ 0.8 GPa of the superconducting dome. This needs to be examined by high pressure magnetization measurements. More interestingly, as the pressure further increases, superconductivity re-emerges, and the $T_c$ keeps increasing up to 47.0 GPa, as shown in Fig. 4d. The effect of field on the resistance transition under 29.3 GPa in Fig. 4e demonstrates it is a superconducting transition. In Fig. 4f, we plot the full temperature – pressure phase diagram, which includes two superconducting domes.

A temperature - pressure ($T_c$ vs $p$) or temperature - doping ($T_c$ vs $x$) superconducting dome has been commonly observed in many unconventional superconductors, including heavy-fermion superconductors, cuprate superconductors, iron-based superconductors, and quasi-two-dimensional organic superconductors[1]. For example, the heavy-fermion superconductor CeCoIn$_5$ manifests a $T_c$ vs $p$ superconducting dome, and the unconventional superconductivity with $d_{x^2-y^2}$ symmetry may result from the antiferromagnetic spin fluctuations[23]. Theoretically, it has been shown that unconventional superconductivity with $d_{xy}$ symmetry can also appear in close proximity to a charge-ordered phase, and the superconductivity is mediated by charge fluctuations[24,25]. This may be the case of the pressure-induced superconductivity in 1$T$-TiSe$_2$, with the superconducting dome appearing around the critical pressure related to the charge-density wave (CDW) meltdown[26]. For Kagome lattice at van Hove filling, previous theoretical calculations found competing electronic orders, including CDW and chiral $d_{x^2-y^2} + id_{xy}$ superconductivity[27,28]. However, proximity-induced spin-triplet superconductivity was claimed in Nb-K$_{1-x}$V$_3$Sb$_5$ devices[29]. Nevertheless, our finding of nodal superconductivity put a strong constrain on the pairing state of the first dome of CsV$_3$Sb$_5$, which should be related to the CDW instability. A second superconducting dome is rare, including the heavy-fermion CeCu$_2$(Si$_{1-x}$Ge$_x$)$_2$ (ref. [30]) and the iron chalcogenides[31]. Since the second dome is away from the CDW order, more experimental and theoretical works are needed to understand the superconductivity of the second dome in CsV$_3$Sb$_5$.

In summary, we investigate the superconducting gap structure of the new V-based superconductor CsV$_3$Sb$_5$ by ultralow-temperature thermal conductivity measurements. The finite $\kappa_0/T$ in zero magnetic field and its rapid field dependence give strong evidences for nodes in the superconducting gap. Further measurements of resistance under pressure reveal two superconducting domes in the temperature - pressure phase diagram. These results suggest unconventional superconductivity in CsV$_3$Sb$_5$. While our finding of nodal superconductivity puts a strong constrain on the pairing state of the first dome, which should be related to the CDW instability, the superconductivity of the second dome may present another exotic pairing state in this ideal Kagome lattice of vanadium.

## Methods

**Sample preparation.** Single crystals of CsV$_3$Sb$_5$ were grown from Cs ingot (purity 99.9%), V powder (purity 99.9%) and Sb grains (purity 99.999%) using the self-flux method[7]. The eutectic mixture of CsSb and CsSb$_2$ is mixed with VSb$_2$ to form a composition with 50 at.% Cs$_x$Sb$_y$ and 50 at.% VSb$_2$ approximately. The mixture was put into an alumina crucible and sealed in a quartz ampoule under partial argon atmosphere. The sealed quartz ampoule was heated to 1273 K for 12 h and soaked there for 24 h. Then it was cooled down to 1173 K at 50 K/h and further to 923 K at a slowly rate. Finally, the ampoule was taken out from the furnace and decanted with a centrifuge to separate CsV$_3$Sb$_5$ single crystals from the flux. CsV$_3$Sb$_5$ single crystals are stable in the air. The X-ray diffraction (XRD) measurement was performed on a typical CsV$_3$Sb$_5$ sample by using an X-ray diffractometer (D8 Advance, Bruker), and determined the largest surface to be the (00$l$) plane.

**DC magnetization measurement.** The DC magnetization measurement was performed down to 1.8 K using a magnetic property measurement system (MPMS, Quantum Design).

**Resistivity and thermal transport measurements.** The sample with dimensions of 1.96 × 0.27 mm$^2$ in the *ab* plane and a thickness of 57 μm along the *c* axis was used for both resistivity and thermal transport measurements at ambient pressure. Four silver wires were attached to the sample with silver paint, which were used for both resistivity and thermal conductivity measurements under ambient pressure, with electrical and heat currents in the *ab* plane. The in-plane resistivity was measured in a $^3$He cryostat. The in-plane thermal conductivity was measured in a dilution refrigerator by using a standard four-wire steady-state method with two $RuO_2$ chip thermometers, calibrated *in situ* against a reference $RuO_2$ thermometer. Magnetic fields were applied perpendicular to the *ab* plane in all measurements. To ensure a homogeneous field distribution in the sample, all fields for resistivity and thermal conductivity measurements were applied at a temperature above $T_c$.

**High pressure measurements.** High pressure resistance of $CsV_3Sb_5$ powder sample (ground from single crystals) was measured in a physical property measurement system (PPMS, Quantum Design) and a $^3$He cryostat by using a diamond anvil cell (DAC). The pressures inside of the DAC were scaled by ruby fluorescence method at room temperature each time before and after the measurement.

## Data availability

The data that support the findings of this study are available from the corresponding author upon reasonable request.

**Acknowledgements**

This work was supported by the Natural Science Foundation of China (Grant No. 12034004), the Ministry of Science and Technology of China (Grant No.: 2016YFA0300503), and the Shanghai Municipal Science and Technology Major Project (Grant No. 2019SHZDZX01). Y. F. Guo was supported by the Major Research Plan of the National Natural Science Foundation of China (No. 92065201) and the Program for Professor of Special Appointment (Shanghai Eastern Scholar). H. C. Lei was supported by National Natural Science Foundation of China (Grant No. 11822412 and 11774423), the Ministry of Science and Technology of China (Grant No. 2018YFE0202600 and 2016YFA0300504), and Beijing Natural Science Foundation (Grant No. Z200005).

## Author Contributions

S.Y.L. conceived the idea and designed the experiments. C.C.Z and L.S.W. performed the DC magnetization and bulk transport measurements with help from J.M.N., Y.Y.H., C.P.T., X.F.Y. was responsible for high-pressure transport experiments. W.X., Z.C.T., Y.F.G., Q.W.Y., Z.J.T., C.S.G., and H.C.L. synthesized the single crystal samples. S.Y.L. wrote the manuscript with comments from all authors. C.C.Z., L.S.W., W.X., and Q.W.Y. contributed equally to this work.

## Competing interests

The authors declare no competing interests.

## Additional information

**Supplementary information** is available for this paper at URL inserted when published

**Correspondence** and requests for materials should be addressed to H.C.L (hlei@ruc.edu.cn;); Y.F.G (guoyf@shanghaitech.edu.cn); X.F.Y. (yangxiaofan@fudan.edu.cn) or S.Y. L. (shiyan_li@fudan.edu.cn)

Figure 1

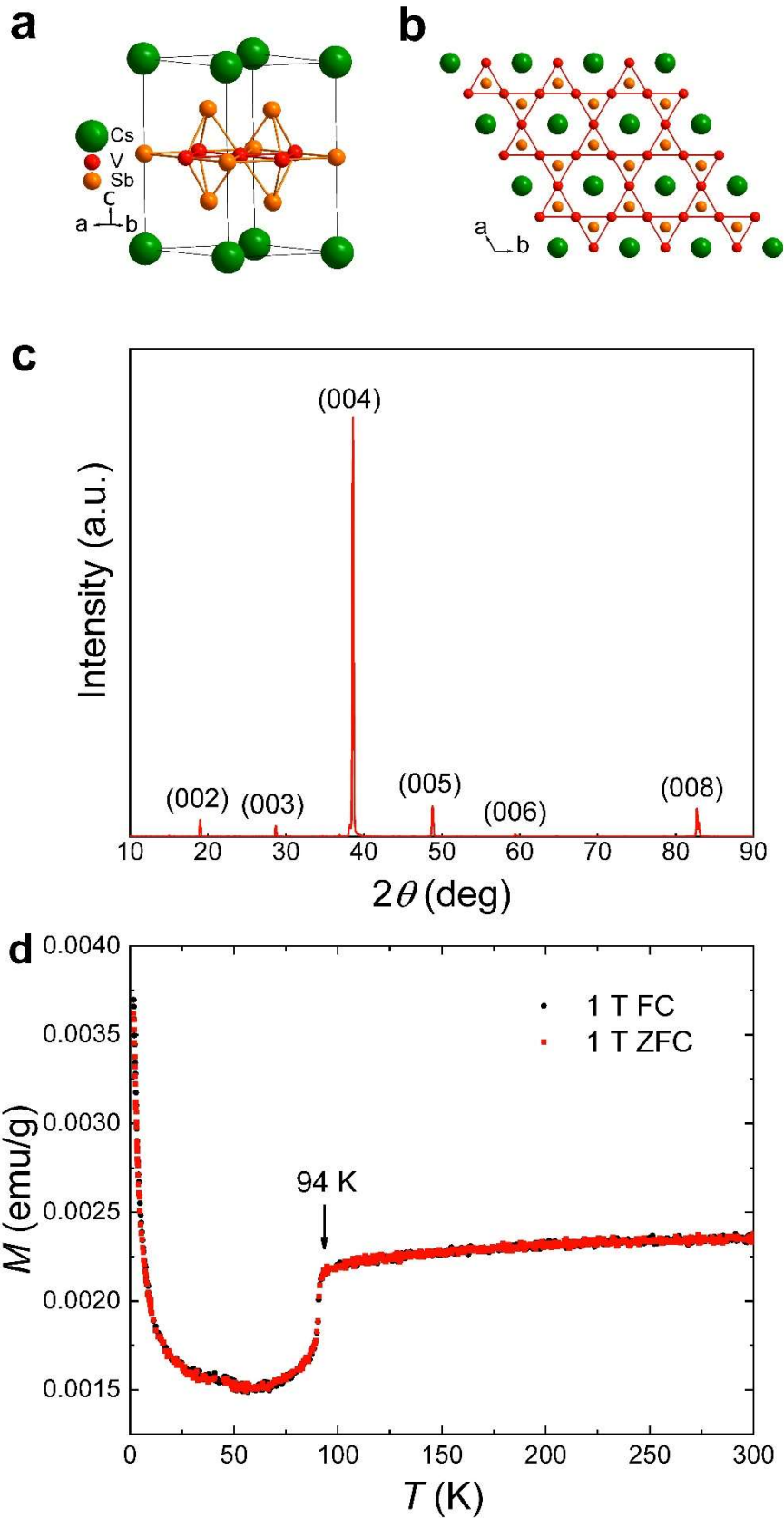

Figure 2

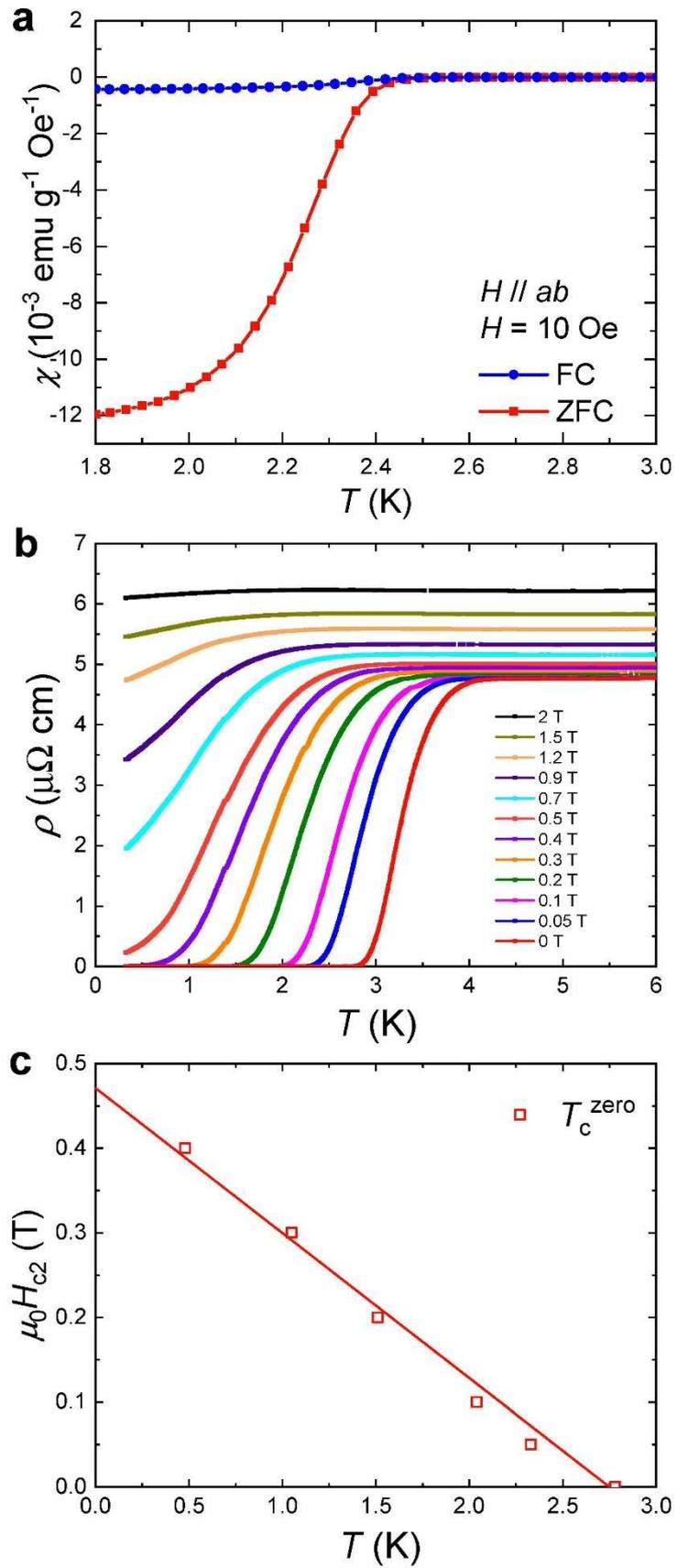

Figure 3

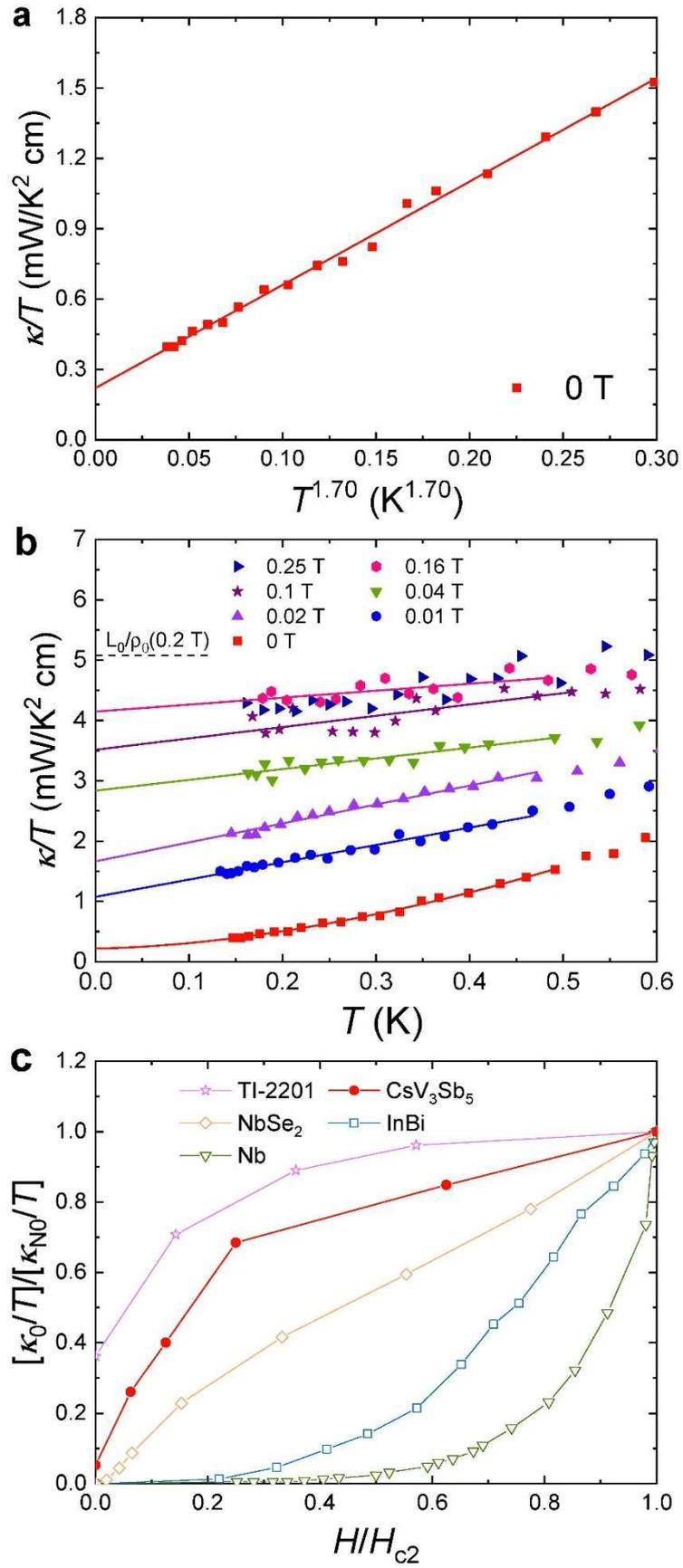

Figure 4

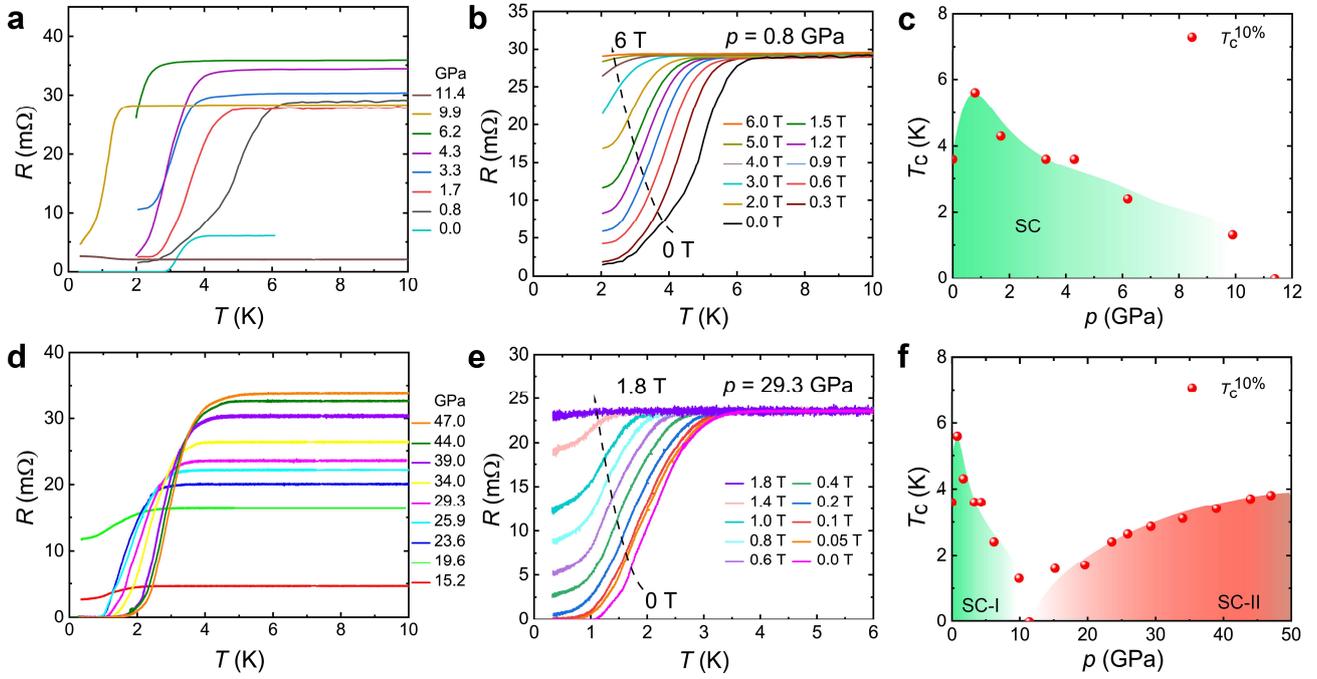

**Figure captions**

**Fig. 1 | Characterization of CsV$_3$Sb$_5$. a** Crystal structure of CsV$_3$Sb$_5$. The Cs, V, Sb atoms are presented as green, red and orange balls, respectively. **b** Top-down view of the crystal structure. The two-dimensional Kagome lattice of vanadium can be clearly seen. **c** Room-temperature X-ray diffraction pattern of the CsV$_3$Sb$_5$ single crystal, showing that the largest natural face is (00$l$) plane. **d** Temperature dependence of the magnetization for CsV$_3$Sb$_5$ single crystal. The arrow denotes a charge density wave order at 94 K.

**Fig. 2 | Superconductivity of CsV$_3$Sb$_5$. a** Low-temperature magnetization of CsV$_3$Sb$_5$ single crystal at $H$ = 10 Oe, with zero-field and field cooling modes, respectively. **b** Low-temperature in-plane resistivity of CsV$_3$Sb$_5$ single crystal in magnetic fields up to 2 T. **c** Temperature dependence of the upper critical field $\mu_0H_{c2}$, extracted from the $T_c^{zero}$ values in panel **b**. The red line is a linear fit to $\mu_0H_{c2}(T)$, and $\mu_0H_{c2}(0) \approx 0.47$ T is roughly estimated.

**Fig. 3 | Thermal conductivity of CsV$_3$Sb$_5$. a** Temperature dependence of the in-plane thermal conductivity for CsV$_3$Sb$_5$ single crystal in zero field. The solid line represents a fit to $\kappa/T = a + bT^{\alpha-1}$ below 0.5 K, which gives the residual linear term $\kappa_0/T \equiv a = 0.22 \pm 0.04$ mW K$^{-2}$ cm$^{-1}$ and $\alpha = 2.70 \pm 0.12$. **b** The thermal conductivity of CsV$_3$Sb$_5$ in magnetic fields up to 0.25 T. The dashed line is the normal-state Wiedemann-Franz law expectation $L_0/\rho_0$ (0.2T), with the Lorenz number 2.45 × 10$^{-8}$ W Ω K$^{-2}$ and $\rho_0(0.2T) = 4.82$ μΩ cm. **c** Normalized residual linear term $\kappa_0/T$ of CsV$_3$Sb$_5$ as a function of $H/H_{c2}$, with bulk $H_{c2} = 0.16$ T. Similar data of the clean $s$-wave superconductor Nb (ref. [22]), the dirty $s$-wave superconducting alloy InBi (ref. [16]), the multiband $s$-wave superconductor NbSe$_2$ (ref. [17]), and an overdoped $d$-wave cuprate superconductor Tl-2201 (ref. [18]) are shown for comparison.

**Fig. 4 | Resistance under pressure and temperature - pressure phase diagram for CsV$_3$Sb$_5$. a** Temperature dependence of resistance for CsV$_3$Sb$_5$ under various pressures up to 11.4 GPa. The curve of 0 GPa is from the CsV$_3$Sb$_5$ single crystal in Fig. 2b. **b** Temperature dependence of resistance for CsV$_3$Sb$_5$ under different magnetic fields at 0.8 GPa. Increasing the magnetic field gradually suppresses the superconducting transition. **c** Temperature - pressure phase diagram up to 11.4 GPa for CsV$_3$Sb$_5$. $T_c$ is determined at the 10% drop of the normal-state resistance. It shows a clear superconducting dome.

**d** Temperature dependence of resistance for CsV$_3$Sb$_5$ under higher pressures up to 47.0 GPa. **e** Temperature dependence of resistance for CsV$_3$Sb$_5$ under different magnetic fields at 29.3 GPa. **f** Temperature - pressure phase diagram up to 47.0 GPa for CsV$_3$Sb$_5$. It shows two superconducting domes.